# Prompt Gamma Timing for range verification with carbon ion irradiation: first experimental measurements and comparison with Geant4 Monte Carlo simulations.

Iram Barbaro Rivas Ortiz*[1], Sahar Ranjbar*[2], Piergiorgio Cerello[1], Emanuele Maria Data[2], Mohammad Fadavi Mazinani[1], Miguel David Fernandez Moreira[1], Veronica Ferrero[2], Simona Giordanengo[1], Felix Mas Milian[1,3], Diango Manuel Montalvan Olivares[1,2], Francesco Pennazio[1], Marco Pullia[4], Roberto Sacchi[1,2], Roberto Cirio[1,2], Simone Savazzi[4], Anna Vignati[1,2], Elisa Fiorina[1]

[1] Istituto Nazionale di Fisica Nucleare Sezione di Torino, Via Pietro Giuria 1, Turin, Piedmont, IT 10125

[2] Università degli Studi di Torino, Via Pietro Giuria 1, Turin, Piedmont, IT 10124

[3] Universidade Estadual de Santa Cruz, Campus Soane Nazaré de Andrade, Rod. Jorge Amado, Km 16 - Salobrinho, Ilhéus, BA, BR 45662-900

[4] Fondazione Centro Nazionale di Adroterapia Oncologica, Research and Development, Via Borloni, 1, Pavia, Lombardia, IT 27100

* Authors equally contributing

## Abstract (300 words)

Prompt Gamma Timing (PGT) is a promising technique for in vivo range verification in particle therapy, exploiting the time-of-flight between primary particles and prompt gamma rays emitted by nuclear interactions. PGT distribution is highly sensitive to beam energy and target density, which, under controlled detector positioning, enables real-time monitoring of particle range, detection of morphological changes, and support for adaptive treatment strategies.

**Objective:** This study investigates for the first time the application of PGT in carbon ion therapy.

**Approach:** Measurements were performed using a dedicated detection system composed of a silicon strip sensor for primary ion timing and a LaBr$_3$(Ce) scintillator read out by a SiPM matrix for secondary radiation. Carbon ion beams with energies of 166.41, 268.86, and 398.84 MeV/u irradiated a homogeneous 30.0 cm PMMA target at CNAO (Pavia, Italy). The secondary radiation detector was positioned iat four off-beam positions to assess the robustness of the PGT technique. Monte Carlo simulations based on the Geant4 were conducted for all configurations to evaluate agreement and predictive capability.

**Main Results:** A quantitative bin-by-bin comparison of experimental and simulated PGT intensities demonstrated strong agreement within the 95% confidence interval, with no incompatible bins at 166.41 MeV/u, at most 1% at 268.86 MeV/u, and up to 8% at 398.84 MeV/u, depending on detector position. Prompt gammas were identified as the dominant contribution to the detected signals, particularly for detector positions upstream with respect to the primary particle beam, minimizing signal contamination from neutrons and charged fragments.

**Significance:** The validated experimental - simulation framework confirms the capability of the proposed PGT system to resolve energy-dependent differences and highlights its potential for detecting clinically relevant changes in the particle beam range, supporting further development toward real-time monitoring in carbon ion therapy.

# 1 Introduction

Proton and carbon ion particle therapy offer precise cancer treatment, resulting in accurate dose delivery and minimal impact on healthy tissue. This precision comes from a narrow energy deposition located close to the end of the particle range in the target, the so-called Bragg peak region (Kraan and Del Guerra 2024). Modern treatment planning systems allow robust and optimized dose delivery considering all the uncertainties that could have an impact on Bragg peak positions at different beam energies; however, factors like patient motion and anatomical changes during treatment can still compromise dose accuracy, highlighting the need for reliable in vivo range verification techniques (Galeone et al. 2025).

Prompt Gamma Timing (PGT) has emerged as a promising technique for real-time range verification in particle therapy. The technique exploits the emission of de-excitation photons due to nuclear interaction between the particle beam and the human tissues during the irradiation. In particular, the PGT approach is based on measuring the time difference between the primary particle crossing the beam monitor and the prompt gamma-ray detection for estimating the particle range with high precision (Krimmer et al. 2018). This time-of-flight information enables real-time monitoring of the treatment quality, making it suitable for adaptive therapy and treatment optimization (Golnik et al. 2014). Further developments investigated the temporal structure of clinical proton beams and their impact on timing-based range verification. In particular, studies by the Dresden group characterized the microbunch time structure for proton pencil beams at clinical facilities and explored signal processing strategies for prompt gamma timing data obtained with cyclotron-based beams (Petzoldt et al. 2016, Werner et al. 2019).

These works contributed to defining the achievable timing performance and practical implementation of PGT in realistic treatment conditions.

Complementary approaches have also been proposed to exploit very high timing resolution in the single-proton counting regime. Marcatili et al. (2020) demonstrated ultra-fast prompt gamma detection combined with single ion tagging, enabling range monitoring based on precise time measurements. This concept was further extended to Prompt Gamma Imaging (PGI) using time-of-flight-based reconstruction methods (Nenoff et al. 2017, Jacquet et al. 2021), and more recently to high-sensitivity Cherenkov detectors optimized for PGT and PGI applications (Jacquet et al. 2023). Recent work has also focused on improving the reconstruction of prompt gamma emission distributions from timing information. Pennazio et al. (2022) introduced an innovative spatio-temporal reconstruction framework based on prompt gamma timing. This approach formulates the problem as an inverse reconstruction using maximum-likelihood methods, enabling the recovery of the spatial distribution of prompt gamma emission, thereby improving robustness and quantitative accuracy of range estimation. The MERLINO-INFN project builds directly on this reconstruction approach, extending it toward the in vivo estimation of proton stopping power. Ferrero et al. (2022) adopt the spatio-temporal reconstruction concept combined with a dedicated multi-detector PGT system. This retrieves energy loss profiles from measured timing distributions and assesses the local stopping power of tissues during irradiation. Prompt gamma emission has also been exploited through approaches based on different observables. In particular, Prompt Gamma Spectroscopy (PGS) focuses on the energy and timing characteristics of prompt photons emitted from nuclear interactions with elements such as carbon and oxygen, enabling range estimation at the level of individual pencil beams (Verburg and Seco

2014). Together with timing-based techniques, these methods form a broader set of prompt gamma techniques currently investigated for in vivo range verification in particle therapy.

The implementation of timing-based monitoring techniques relies critically on the performance of both the beam monitoring and prompt gamma detection systems. Achieving single-particle tagging with high temporal resolution remains a challenging task, and several detector technologies have been investigated for this purpose. Fast silicon sensors provide excellent timing performance and compact integration in beamline environments, while alternative solutions such as CVD diamond detectors have demonstrated promising characteristics in terms of timing resolution and radiation hardness for beam tagging applications in hadron therapy (Curtoni et al. 2021). Similarly, the choice of scintillation detector plays a key role in prompt gamma timing measurements. Several scintillator materials have been investigated for prompt gamma detection in particle therapy, including $BaF_2$, LYSO, and $CeBr_3$, each offering different trade-offs between timing performance, detection efficiency, and energy resolution (Lecoq 2016, Krimmer et al. 2018). $BaF_2$ provides an extremely fast scintillation component that is advantageous for timing applications, although its relatively low light yield and the presence of a slower component may limit energy resolution. LYSO crystals offer high density and excellent detection efficiency but exhibit slower scintillation decay times, which can reduce timing performance. Lanthanum- and cerium-based bromide scintillators have attracted particular interest for prompt gamma detection due to their combination of high light yield, fast decay time, and excellent energy resolution. In particular, $LaBr_3$ and $CeBr_3$ scintillators coupled to Silicon photomultipliers (SiPMs) share a favourable compromise between timing performance making them well suited for fast timing applications (Fraile et al. 2013, Vedia et al. 2017). $CeBr_3$ presents the additional advantage of significantly lower intrinsic radioactivity compared with

LaBr$_3$, which can be beneficial in low background measurements. Nevertheless, both materials have been successfully employed in prompt gamma detection systems, and their performance characteristics make them attractive candidates for timing-based monitoring approaches in particle therapy.

Monte Carlo (MC) simulations are widely used in particle therapy due to their ability to model particle transport and nuclear interactions in complex geometries (Muraro et al. 2020, Freitas et al. 2025, Lai et al. 2025). They support a broad range of applications, including dose calculations (Lysakovski et al. 2021), radiobiological modeling (Rucinski et al. 2021), and imaging system development (Eo et al. 2025). Monte Carlo methods are also essential for range verification, as they can accurately model the transport of primary particles and the production of secondaries within the target and surrounding tissues (Dedes et al. 2014, Kraan 2015, Ferrero et al. 2022, Everaere et al. 2024), a feature that is particularly important for techniques such as PGT, where carbon ion beams offer higher biological effectiveness than protons but produce more nuclear fragments and secondary particles, increasing range uncertainty and modeling complexity.

Recently, our research group have started to explore PGT measurements with carbon ion beams, as reported in Ranjbar et al. (2024), highlighting PGT performance in different target geometries using 398.84 MeV/u carbon ions and introducing a novel compatibility analysis for meaningful comparison across different target configurations. Therefore, the present study reports PGT measurements performed with carbon ion beams on a homogeneous phantom target and different beam energies, along with extensive comparison analysis to MC simulations. A dedicated PGT setup was built, combining a silicon beam monitor providing single-particle timing and a secondary radiation detector with high temporal performance, placed around a thick PMMA target at four different positions with respect to the isocentre. Three different carbon ion energies were investigated. The following sections describe the experimental and MC setup used to measure PGT distributions and the compatibility analysis carried out to compare experimental and simulated results. Leveraging MC truth information, the contributions of different secondary particles to the recorded signals were also identified, providing insight into the underlying features of the observed PGT distributions. The primary goal of this work is to establish the methodology of PGT for carbon ions and demonstrate the feasibility of the technique, forming a foundation for future studies on range sensitivity and more complex target configurations.

## 2 Material and Methods

### 2.1 The PGT system experimental layout

The PGT system includes a beam sensor for monitoring the incident primary particle beam, and a secondary radiation detector to record the arrival time of prompt gammas. The beam monitor consists of an 8-strip silicon sensor, each strip having an active area of 4 mm × 0.55 mm, resulting in a total active area of 17.6 mm$^2$ (Olivares et al. 2025). Prompt Gamma (PG) detection is performed using a LaBr$_3$(Ce) scintillator (cylindrical shape, 19.05 mm radius, 38.10 mm thickness) read out by a 5 x 5 Silicon PhotoMultiplier (SiPM) RGB-HD matrix (24 x 24 mm$^2$ active area) (Bartosik et al. 2025) (Fig. 1a). The beam monitor achieves a time resolution of approximately 30 ps, while the secondary radiation detector shows a resolution of about 200 ps, as assessed through measurements using a $^{60}$Co radioactive source. Assuming independent contributions, the theoretical timing resolution of the system is given by the quadratic sum of both terms, resulting in approximately 202 ps,, which is sufficient to acquire meaningful PGT distributions sensitive to density variations on the order of a few millimeters in depth. The DT5742 CAEN digitizer was used for signal readout at a 2.5 GHz sampling rate, with a dead time of approximately 110 µs per event due to conversion. An energy window of about 1.0 to 10.0 MeV was applied to secondary radiation signals to discard most of the unwanted events (e.g., charged fragments).

The measurements were carried out at the National Centre of Oncological Hadrontherapy - CNAO (Pavia, Italy). Carbon ion beams with 166.41, 268.86, and 398.84 MeV/u energies irradiated a homogeneous 30.0 cm thick polymethyl methacrylate (PMMA) phantom target. The beam sensor was placed on the beam trajectory, 42.5 cm upstream of the room isocentre. The secondary detector was positioned at a distance of 20.1 cm from the isocentre at different polar angles with respect to the beam axis: 25.91° (position D), 51.43° (C), 90.00° (B), and 141.43° (A). Each experimental configuration was acquired for approximately 5 minutes at a beam intensity of about $5\times10^6$ pps ($2\times10^8$ pps instantaneous rate), corresponding to roughly 1/100 of the typical clinical rate used during treatment delivery. At the investigated energies, the carbon ions travel with velocities of about 15.6-21.2 cm/ns, corresponding to 1.4-1.9 ns of minimum times required by carbon ions to be stopped inside the 30 cm PMMA target, depending on the initial beam energy, with slightly longer times expected when accounting for energy loss during transport. The temporal structure of the beam is determined by the accelerator radiofrequency system, resulting in a period ranging between 500 ns for the lowest and 360 ns for the highest beam energy. As a consequence, the instantaneous particle rate within individual micropulses is significantly higher than the average rate over longer time scales between beam micropulses. The subclinical rate used in the present measurements was chosen to limit event pile-up while preserving the relevant timing characteristics of the beam structure.

At the isocentre, the carbon ion beams exhibit a two-dimensional Gaussian spatial profile with Full Width Half Maximum (FWHM) values of 0.69 cm, 0.52 cm and 0.45 cm for 166.41 MeV/u, 268.86 MeV/u, and 398.84 MeV/u, respectively, as assessed by the CNAO commissioning measurements in air (Mirandola et al. 2015). Considering the beam dimensions and the active area of the beam monitor, the sensor intercepts only a fraction of the primary ions. A simple geometrical estimate indicates that the monitor intercepts roughly 30-75% of the beam intensity depending on the beam energy and transverse spread. While this fraction is sufficient for the present proof-of-concept study, a clinical implementation would require a larger active area in order to tag a higher fraction of the primary ions.

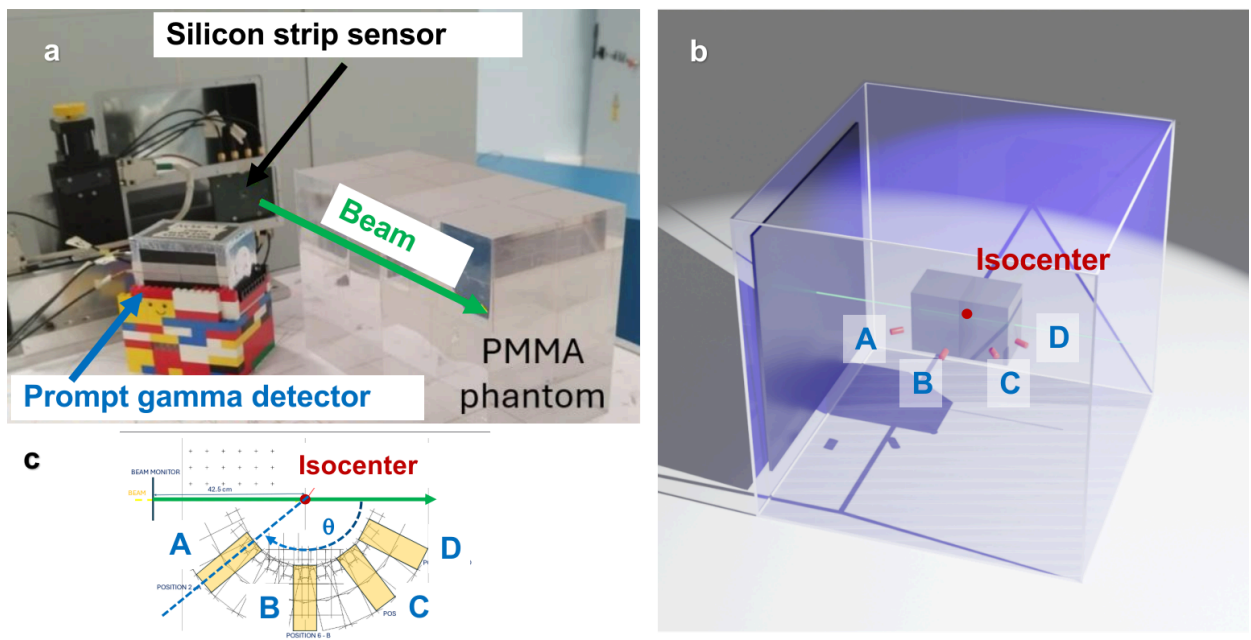

Figure 1 - Experimental (a) and MC (b) configuration for beam test at CNAO (September 2024). Carbon ion beam (green line) irradiating a 30.0 cm-thick PMMA target. LaBr$_3$(Ce) scintillators (blue tags) were positioned at a distance of 20.1 cm from the geometry isocenter (red dot), at different polar angles with respect to the beam axis (c): 25.71° (D), 51.43° (C), 90.00° (B), and 141.43° (A).

## 2.2 The Monte Carlo simulation tool

MC simulations were performed using the Geant4 MC toolkit, version 11.2.2. Hadron-nucleus interactions were modeled using the Binary Intranuclear Cascade (BIC) model, coupled with the Precompound model in the equilibrium phase, contained in the reference physics list QGSP_BIC_HP (Everaere et al. 2024, Freitas et al. 2025, Lai et al. 2025). For nucleus-nucleus interactions, the Quantum Molecular Dynamics (QMD) model was used (Dedes et al. 2014). In both the BIC and QMD models, the default Geant4 de-excitation model was used, including fragmentation, evaporation, Fermi break-up, and gamma emission, depending on the mass and excitation energy of the nucleus. For electromagnetic processes, the predefined Geant4 Standard Electromagnetic Physics Option 4 model was employed. All particle types below a threshold corresponding to a minimum energy of 0.01 MeV were neglected. The carbon ion beam was modeled as a two-dimensional Gaussian spatial profile, using the FWHM values corresponding to each beam energy as reported in the previous section.

To simulate the secondary radiation response, a custom code was integrated into the Geant4 toolkit, accounting for the time and cumulative energy deposition for each event scoring at each secondary radiation detector position. Cumulative energy deposition was linked to each secondary particle entering the crystals and updated for each energy deposition event, including descendant cascade contributions. The trigger particle type, kinetic energy, production position, total energy deposition, and starting time of the cascade process were recorded for each secondary particle. A time threshold of 50 ns was used to stop the tracking process, excluding longer radioactive decay channels. The beam monitor detector was not simulated, and the delivery time of the primaries was set equal to 0 ns. The energy deposition data were employed to filter events within the 1-10.0 MeV energy window in agreement with the experimental data analysis. Time data was used to construct PGT distributions, while particle type information was used to evaluate the contributions of different particles. Production positions of secondary particles were analyzed to further investigate the spatial distribution of photon production as a function of carbon ion energy, while their kinetic energy was used to generate prompt gamma energy spectra.

*2.3 Signal timing and combinatorial background*

The acquisition is triggered by signals from the secondary radiation detector exceeding a predefined voltage threshold ($\geq 22$ mV, corresponding to $\sim 1$ MeV). Because of the beam delivery time profile, multiple carbon ion signals may be recorded for a single secondary event that triggers the acquisition. Figure 2A shows a representative digitizer snapshot containing three carbon ion waveforms (one from strip 1 and two from strip 2) together with the corresponding secondary radiation signal. The arrival time of the secondary signal was extracted from the rising edge of the waveform using a linear fit, with the time defined at 10% of the maximum amplitude. For carbon ion signals, the timing was obtained from the peak position of a Gaussian fit. From the snapshot data, three possible time-difference pairings can be formed. However, because the silicon sensor is smaller than the beam FWHM, true coincidences may be missed. Consequently, no more than one (and possibly none) of the pairings corresponds to a true coincidence, while the remaining pairings arise from uncorrelated hits and represent random coincidences. The accumulation of these random pairings per trigger event leads to a combinatorial background, which is reflected in the PGT distribution.

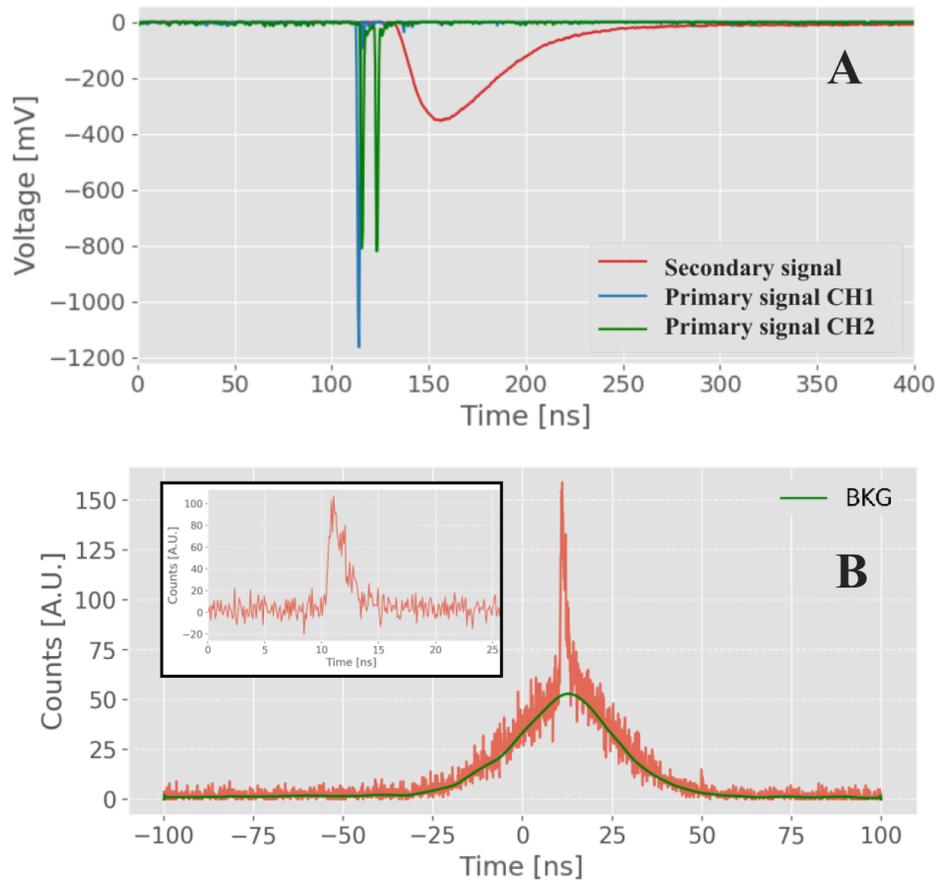

Figure 2 - (A) Example of digitizer snapshot containing the waveforms of carbon ion signals (blue - channel 1, green - channel 2), and the secondary signal (red) for a single event. (B) PGT distribution (red) with combinatorial background (green) from the overall data acquisition. Experimental PGT distribution after combinatorial background removal (inner plot).

Figure 2B shows an example of PGT distribution obtained with a fixed bin width of 100 ps. The accumulation of random time coincidences during data acquisition produces a significant combinatorial background (green curve), which must be subtracted. True coincidences, corresponding to the correct association between the primary particle and the secondary radiation, appear on top of the distribution, forming a sharp peak above the combinatorial background. The inner plot in Figure 2A illustrates the corresponding PGT distribution after combinatorial background removal using the Statistical-sensitive Non-linear Iterative Peak-clipping (SNIP) algorithm (Ryan et al. 1988) with 20 iterations.

*2.4 Dataset sampling size and normalization*

The CNAO Dose Delivery System (DDS) and the acquisition setup used in this work do not allow a direct determination of the number of incident ions associated with each dataset. This limitation arises from several factors: (a) the DDS is not sensitive enough at subclinical rates to deliver the same number of primaries across different acquisitions; (b) the data acquisition is triggered by the detection of secondary radiation rather than the primary ions passing through the beam monitor; (c) the acquisition dead time is not precisely known; and (d) the beam monitor samples only a fraction of the incident ions because the beam FWHM is larger that the active width of the detector. A normalization procedure based on MC simulations was therefore implemented to define the sampling size of the experimental datasets for each detector-energy configuration. The goal of this procedure is double-fold. Firstly, normalization allows the construction of reliable PGT time distributions while preserving both the average shape and the statistical fluctuations in each time bin. Second, it provides an estimate of the number of primary ions delivered for each configuration, thereby compensating for the experimental limitations related to triggering and acquisition dead time. As a result, the normalized PGT distributions enable a more meaningful assessment of the technique under clinical-like scenarios.

To determine the appropriate sampling size for the experimental datasets, MC simulations were performed using $1\times10^6$ primary histories and repeated 100 times. These simulations provided estimates of the average number of secondary particles depositing energy in the detector for each detector position-carbon ion energy configuration. Table 1 summarizes the experimental statistics recorded for each configuration together with the corresponding MC predictions. The table reports the number of experimental time differences between the beam monitor and the secondary radiation detector, the number of experimental true coincidences obtained after combinatorial background subtraction, and the expected number of true coincidences predicted by the MC simulations.

Table 1 - Experimental acquisition statistics and MC predictions used for dataset normalization

| Beam Energy | Detector position | Time differences | True Coincidences | MC Predictions | Ratio (exp/MC) |
|---|---|---|---|---|---|
| 398.84 MeV/u | A | 91200±300* | 4760±70 | 950±30 | 5.01 |
|  | B | 50400±200 | 2070±50 | 1220±40 | 1.70 |
|  | C | 126300±400 | 9950±100 | 2090±40 | 4.76 |
|  | **D** | **55000±200** | **6010±80** | **3770±60** | **1.59** |
| 268.86 MeV/u | A | 1330400±1200 | 33890±180 | 810±20 | 41.9 |
|  | B | 1003400±1000 | 26740±160 | 830±30 | 32.2 |
|  | C | 465100±700 | 17990±130 | 1120±40 | 16.1 |
|  | D | 239900±500 | 11300±110 | 1640±40 | 6.89 |
| 166.41 MeV/u | A | 937600±1000 | 23350±150 | 520±20 | 44.9 |
|  | B | 883400±900 | 19490±140 | 370±20 | 52.7 |
|  | C | 634000±800 | 17360±130 | 400±20 | 43.4 |
|  | D | 596600±800 | 19320±140 | 420±20 | 46.0 |

* Uncertainties correspond to the square root of the number of counts assuming Poisson statistics.

The normalization was determined from the configuration corresponding to detector position D and 398.84 MeV/u beam energy, for which the ratio between the measured number of true coincidences and the MC prediction equals 1.59, representing the lowest experimental-to-MC proportion among all configurations. This conservative choice avoids oversampling experimental datasets with lower statistics and ensure a consistent normalization across all measurements. The corresponding scaling factor therefore implies an effective number of approximately $1.59 \times 10^6$ carbon ions associated with the experimental dataset. For reference, in clinical carbon ion therapy a single pencil beam spot typically delivers on the order of $1\text{-}2 \times 10^5$ primary ions, depending on the treatment plan and beam energy. The number of primaries considered in the present measurements therefore corresponds to approximately 6-9 clinical treatment spots.

Table 2 reports the sampling sizes used for each detector position and beam energy configuration. These values correspond to the number of time differences randomly selected from the experimental datasets in order to reproduce the MC prediction scaled by the normalization factor S = 1.59. In practice, the sampling size $N_{sample}$ is determined by selecting a subset of the experimental time difference values such that the expected number of true coincidences within the sampled dataset matches the MC prediction scaled by the normalization factor. This condition can be expressed as

$$N_{sample} = \frac{S \cdot N_{true}^{MC}}{f_{exp}}, \qquad (1)$$

where $N_{true}^{MC}$ is the MC-predicted number of true coincidences. The experimental fraction $f_{exp}$ is obtained from the ratio between the experimental true coincidences and the total number of time difference values reported in Table 1. By sampling the experimental datasets according to the sizes reported in Table 2, the same effective number of primaries is shared across all detector positions and beam energies. Consequently, the sampling sizes rely on MC-based normalization rather than on a direct experimental count of incident ions. This procedure preserves both the expected prompt gamma signal fraction and the statistical fluctuation of the measured time distributions.

Table 2 - Sampling sizes obtained from experimental TOF datasets after MC-based normalization. These values correspond to subsets of the experimental time-difference datasets selected according to Eq. (1) so that the MC-predicted signal fraction scaled by S = 1.59 is preserved, corresponding to an effective number of $1.59 \times 10^6 \pm 3.23 \times 10^4$ carbon ions.

| Beam Energy | Detector Position A | Detector Position B | Detector Position C | Detector Position D |
|---|---|---|---|---|
| 398.84 MeV/u | 29000±1000* | 47000±1700 | 42300±900 | 55000±1100 |
| 268.86 MeV/u | 55000±1500 | 50000±1900 | 46000±1600 | 55000±1400 |
| 166.41 MeV/u | 33000±1600 | 27000±1400 | 23000±1200 | 21000±1000 |

* The reported uncertainties reflect statistical fluctuations propagated from the MC estimates

## *2.5 Timing parameters for MC comparison*

A meaningful comparison between experimental and simulated PGT time distributions requires accounting for distortions introduced by the experimental detection chain. In particular, the overall time resolution and the transmission delay of the data acquisition system affect both temporal alignment and the shape of the measured distributions. The transmission delay does not carry physical information, as it mainly reflects latencies introduced by the electronic chain and the data acquisition system. In contrast, the system time resolution is a key parameter for the interpretation of PGT measurements and is typically determined through dedicated detector timing characterization. In the present analysis, these parameters were estimated only to determine the time shift and temporal smearing that must be applied to the simulated data, enabling a realistic modeling of the detector response. Therefore, the estimation procedure described here is not intended to provide an independent measurement of the intrinsic detector timing performance. Rather, simulations are used as a physically consistent reference, benefiting from the availability of MC ground truth information on particle type and emission timing.

Two main aspects were addressed: (1) removal of combinatorial background in the experimental data, and (2) estimation of the system time resolution and transmission delay required for the MC comparison. These two correlated parameters were determined using the response surface technique (Labrada et al. 2018), based on a second-order regression model, allowing the identification of the parameter combination that best reproduces the measured PGT distributions. A two-dimensional grid was defined around initial parameter estimates, with a step size of 10 ps in each dimension. For each detector position and carbon ion energy configuration, an experimental PGT distribution was constructed as the reference distribution. Specifically, 2000 bootstrap datasets were generated by sampling with replacement from the original experimental data using the previously determined sampling size. For each subset, time distributions were computed with a bin width of 100 ps and corrected for combinatorial background. The experimental reference distribution was then defined as the median (50th percentile) of these bootstrap distributions. At each grid node, MC data were processed analogously:

(a) 200 MC bootstrap subsets were generated from the simulated dataset (sampling with replacement).

(b) For each subset, time values were smeared according to the time resolution associated with the grid node and shifted according to the corresponding transmission delay.

(c) MC time distributions were then constructed using the same bin width and normalized to the area under the PGT signal of the experimental reference distribution.

The agreement between the experimental reference distribution and the MC distributions was quantified at each grid node using the Pearson correlation coefficient. This procedure was repeated independently for each detector position and carbon ion energy combination, resulting in 2400 Pearson coefficients per grid point. These values populate a three-dimensional parameter space (time delay, time resolution, Pearson coefficient), from which a response surface was constructed using the mathematical model:

$$R = \alpha_1 [t_{del}]^2 + \alpha_2 [t_{res}]^2 + \alpha_3 [t_{del}][t_{res}] + \alpha_4 [t_{del}] + \alpha_5 [t_{res}] + \alpha_6, \tag{2}$$

where $t_{del}$ and $t_{res}$ represent the transmission delay and time resolution, respectively. $R$ is the Pearson correlation coefficient, and $\alpha_i, i = 1{:}6,$ are model parameters. The optimal time resolution and transmission delay were obtained from the maximum of the response surface and applied consistently across all detector positions and beam energy combinations.

After determining the optimal parameters, 2000 bootstrap subsets were again generated from both the experimental and MC datasets. The experimental subsets were corrected using the estimated transmission delay, and the MC subsets were smeared using the estimated time resolution expressed in terms of Root Mean Square (RMS) values. From the resulting bootstrap distributions, 95% Confidence Intervals (CIs) were derived. This step was performed to quantify the impact of statistical fluctuations when comparing experimental and MC PGT distributions.

*2.6 Compatibility analysis*

As established in the previous sections, a bootstrapping procedure was employed to generate statistically comparable datasets across different detector positions and beam energies. This approach effectively normalized both experimental and MC datasets to an equivalent number of primary particles, thereby enabling a meaningful comparison across different acquisition configurations. Following this normalization, a bin-by-bin compatibility analysis was performed to assess the agreement between MC and experimental PGT distributions. Unlike the global correlation-based optimization described in the previous section, this analysis examines each time bin individually. For each bin, the probability (p-value) that a given bin of the simulated PGT distribution falls within the 95% CI of the corresponding bin of the experimental distribution was calculated as

$$pvalue_i = \frac{[counts\ 95\%\ CI_i]}{[total\ counts]}, \qquad (3)$$

thereby providing a localized assessment of statistical compatibility between the two distributions.

## 3 Results and Discussions

### 3.1 Timing parameters for MC comparison

After the regression analysis described in the previous section, a mathematical model describing the correlation between MC and experimental PGT distributions as a function of transmission delay and time resolution was obtained. Table 3 summarizes the parameters of the regression analysis of the fitting method across all detector positions and carbon ion energy combinations.

All coefficients of the mathematical model were found to be statistically significant (p-value < 0.05).

Table 3 - Regression analysis across detector position - carbon ion energy combinations.

| Parameter | Estimate | Std. Error | t-value | p-value |
|---|---|---|---|---|
| $\alpha_1$ | -1.06 | 0.027 | -38.87 | <0.001 |
| $\alpha_2$ | -0.51 | 0.031 | -16.34 | <0.001 |
| $\alpha_3$ | -0.25 | 0.029 | -8.55 | <0.001 |
| $\alpha_4$ | 19.69 | 0.504 | 39.07 | <0.001 |
| $\alpha_5$ | 2.53 | 0.265 | 9.55 | <0.001 |
| $\alpha_6$ | -90.23 | 2.321 | -38.88 | <0.001 |

Figure 3 shows the response surface of the mathematical model. The maximum correlation between simulated and experimental data was obtained at a time resolution of 0.26 ns (RMS) and a transmission delay of 9.23 ns. The transmission delay does not have physical significance and is used only to align the simulated and experimental time distributions. The time resolution parameter represents the effective temporal smearing required to reproduce the detector response in the comparison with MC simulations. This value is slightly larger than the theoretical resolution possible due to additional timing contributions arising from electronics, synchronization, and bem-related fluctuations. For the subsequent data analysis and PGT distribution comparisons, these parameters were applied to the simulated data, resulting in a time shift (due to transmission delay) and a temporal smearing (related to the effective time resolution of the experimental setup).

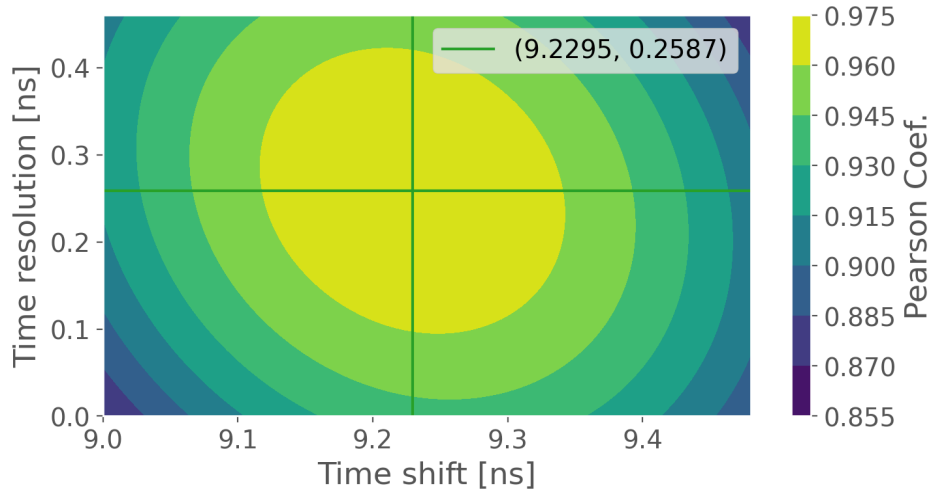

Figure 3 - Response surface of the regression model. Maximum correlation located at 9.23 ns of transmission delay and 0.26 ns (RMS) of time resolution (green lines).

*3.2 Experimental PGT distributions*

Following the PGT distribution calculation described in the previous sections and applying the bootstrapping approach to evaluate the statistical fluctuations, the final results for each detector configuration and beam energy were obtained. Figure 4 shows the experimental PGT distributions for each detector position and carbon ion energy. Each subplot corresponds to a single position and includes the distributions for beam energies of 166.41 MeV/u (red), 268.86 MeV/u (green), and 398.84 MeV/u (blue). The shaded regions represent the 95% CIs. As expected, the PGT distribution shape and statistics strongly depend on the secondary radiation detector position (i.e., A, B, C, D). For a given detector position, both the distribution width and the maximum height vary with beam energy, indicating correlations with carbon ion energy and beam range.

More pronounced energy-dependent differences in distribution width are observed at detector positions A and B (upstream configurations). This behavior can be interpreted considering that the measured PGT corresponds to the sum of two contributions: the time required for the primary ion to reach the gamma emission point and the time required for the emitted gamma to travel from that location to the detector. In upstream configurations, both contributions increase with increasing emission depth (i.e., increasing primary particle range), leading to cumulative timing variations and therefore enhanced sensitivity of the distribution width to beam energy. In contrast, for detector positions C and D (downstream configurations), the ion travel time increases with depth while the gamma travel time decreases because the emission point is closer to the detector. These opposing trends partially compensate each other, reducing the net timing variation associated with changes in beam energy. Consequently, energy-dependent differences in distribution width are less pronounced at positions C and D. Expanding the analysis to include detector positions provides several advantages. It increases the statistical precision of measured distributions, improves the overall reconstruction of the emission profile, and allows a more comprehensive mapping of energy-dependent effects across the full detector array. Incorporating positions at intermediate angles and distances can further disentangle contributions from ion and gamma travel times, enhancing the robustness of timing analysis.

Moreover, larger variations in maximum height are observed at detector positions C and D. This behavior likely reflects a combined effect, including an increased contribution from secondary radiation. At detector positions C and D, a higher amount of secondary radiation signals is expected due to the detection of secondary fragments produced by nuclear interactions between carbon ions and target nuclei, characterized by a strong anisotropic angular distribution. It is important to note that the applied 1-10 MeV energy cut does not preserve the system to be sensitive exclusively to photons (e.g., neutrons and charged fragments). In addition, higher beam energies produce faster secondary fragments, which may reach the secondary radiation detector close in time to the prompt gamma signal, resulting in a cumulative peak with increased PGT maximum height in the time distributions.

In all configurations, a systematic shift in the rising edge of the distributions is also observed. This effect is primarily associated with the time required for the particle beam to reach the target, which decreases with increasing carbon ion energy.

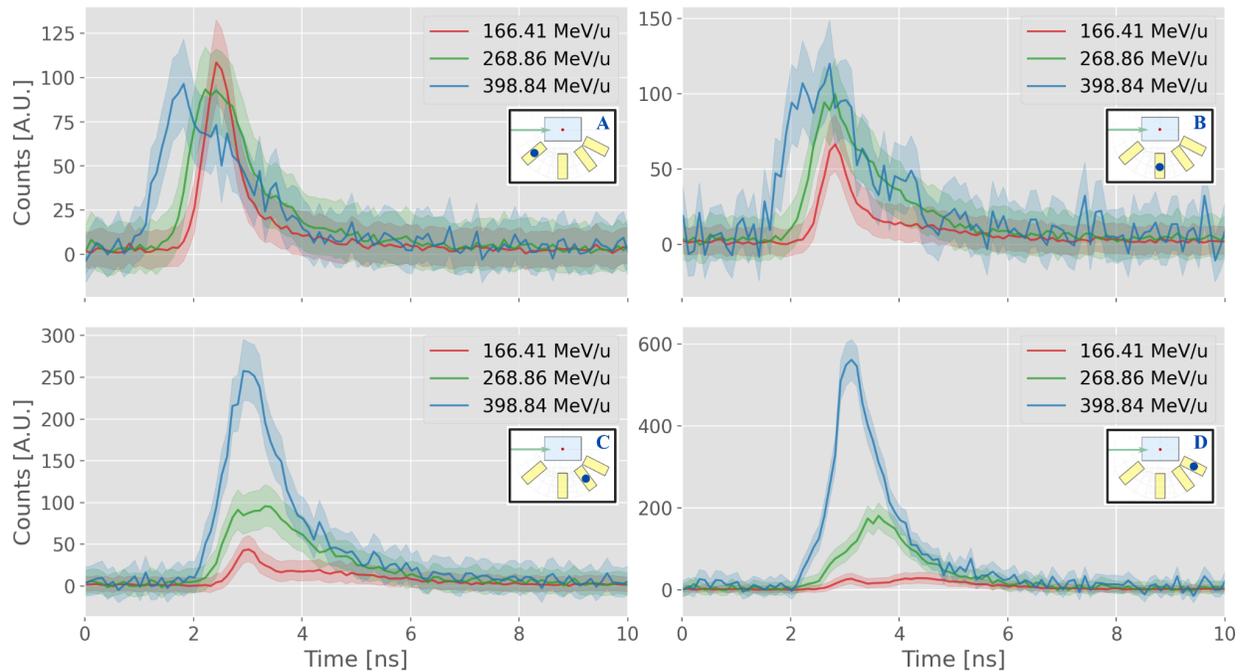

Figure 4 - Experimental PGT distributions as a function of detector position and carbon ion energy. Detector positions A, B, C, D, starting from top left up to bottom right. Results obtained considering the same statistics in all acquisitions (i.e., corresponding to $1.6 \times 10^6$ carbon ions delivered in the MC simulations).

### 3.3 Comparison between MC and experimental distributions

Figure 5 shows the resulting time distributions from both experimental and MC simulations at each detector position for a 166.41 MeV/u carbon ion beam. The corresponding distributions for carbon ion beams of 268.86 MeV/u and 398.84 MeV/u are provided in the Supplementary Materials. Each subplot displays the experimental (red) and MC-simulated (blue) distributions with shaded regions indicating the 95% CIs. The bin-by-bin compatibility test is shown in the top panel of each subplot, where green shades indicate high compatibility and red highlights indicate significant differences (p-value < 0.05). The quantitative compatibility results for all energies and detector positions are summarized as follows. For 166.41 MeV/u, no incompatible bins are observed across any detector positions, with A: 0/100 bins (min p-value=0.16), B: 0/100 bins (min p-value = 0.60), C: 0/100 bins (min p-value = 0.67), and D: 0/100 bins (min p-value = 0.93). For 268.86 MeV/u, only 1 out of 100 bins is incompatible at position C, while the other positions show full compatibility, with A: 0/100 bins (min p-value = 0.70), B: 0/100 bins (min p-value = 0.28), C: 1/100 bins (min p-value = 0.03), and D: 0/100 bins (min p-value = 0.07). For 398.84 MeV/u, the number of incompatible bins varies with detector positions, with A: 0/100 bins (min p-values = 0.65), B: 6/100 bins (min p-value < 0.01), C: 1/100 bins (min p-value = 0.02), and D: 8/100 bins (min p-value < 0.01).

Overall, MC simulations accurately reproduce the observed experimental trends and also fluctuations in the signal region. For irradiation with carbon ions at 166.41 MeV/u, as the detector is moved forward to positions C and D, additional signals become increasingly relevant, forming a second peak in the time distributions. This effect is mainly attributed to the detection of secondary fragments originating from nuclear interactions and produced in a highly anisotropic angular distribution in the forward direction (Nandy 2021). At higher energies (see also the Supplementary Material), the production of secondary fragments occurs, but it is more difficult to observe as a separated peak due to the higher kinetic energy of the secondary fragments, which makes the corresponding signals more likely to overlap with the prompt gamma signals.

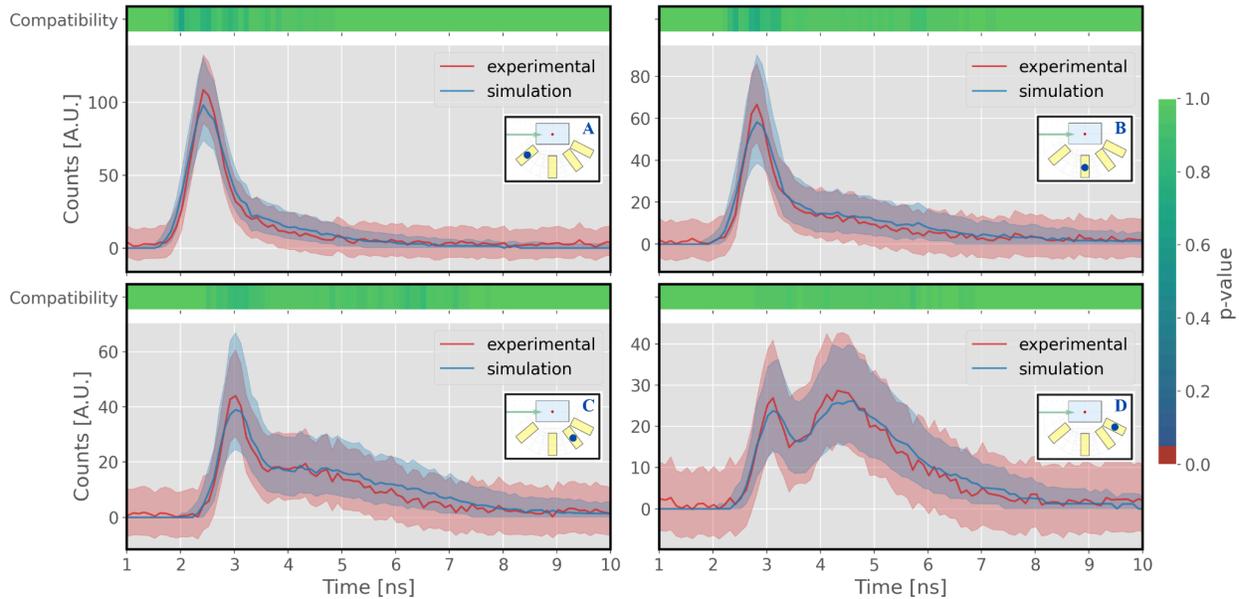

Figure 5 - Experimental and MC simulated time distributions for a 166.41 MeV/u carbon ion beam at each detector position (shown in the box in each subplot). For each position of the secondary radiation detector, compatibility test (top panel) and PGT distribution histograms with 95% CIs (main panel) are shown.

*3.4 Particle contributions*

To further explain the different PGT distribution shapes, the contributions from different secondary particle types are investigated for each detector position, as shown in Figure 6 for the 166.41 MeV/u carbon ion acquisitions. In each subplot, the total distribution is split into contributions from prompt gammas (red), neutrons (blue), protons (purple), electrons (gray), and positrons (yellow), as scored by the MC simulation tool. For the A and B positions, the primary contribution to the overall signal comes from gamma rays, mostly corresponding to the main single peak visible in the PGT distributions. In contrast, at detector positions C and D, neutron contributions become more significant, leading to a slower secondary component that is visible in the PGT distributions. Indeed, at position D, the neutron component is particularly prominent and is identified as the primary contributor to the second peak observed in both experimental and MC results. Protons and electrons contribute marginally to the total signals, though their presence remains detectable. Recent characterization studies of the $LaBr_3(Ce)$ detector crystals support the MC findings regarding the capability of the PGT system to detect fast neutrons (Ranga et al. 2024). Furthermore, Cazzaniga et al. 2015 describe the interaction channels through which neutrons can be detected in $LaBr_3(Ce)$ crystals.

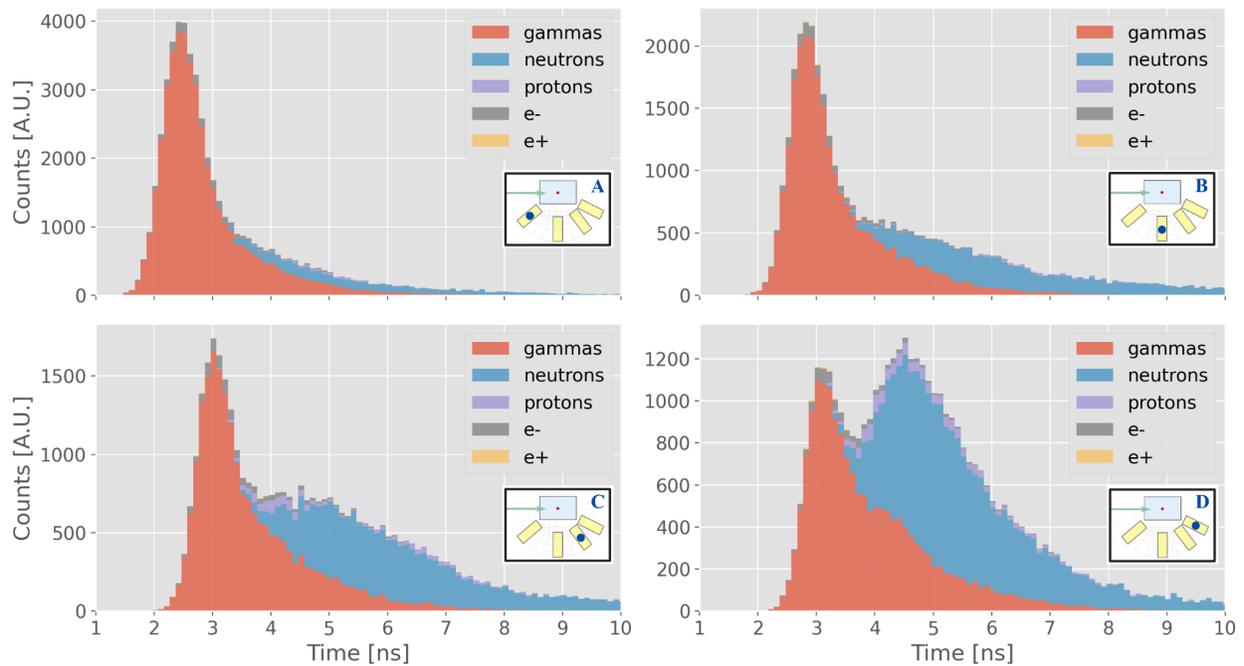

Figure 6 - Particle contribution to the PGT distributions for a 166.41 MeV/u carbon ion beam at detector position A (top left), B (top right), C (bottom left), and D (bottom right), obtained from MC simulations using $10^8$ carbon ions.

Particle contributions to the overall signals for 268.86 MeV/u (Figure 11, Supplementary Materials) and 398.84 MeV/u (Figure 12, Suppelementary Materials) carbon ion beams corroborate the stacking phenomena observed in experimental distributions, mainly due to secondary neutrons and protons, and explain the longer tail behaviour observed in experimental data at detector positions B, C, and D. This behavior has been reported in previous studies investigating secondary proton emission in ion therapy beams (Henriquet et al. 2012). The energy selection applied in this work (1-10 MeV deposited energy) substantially reduces the contribution from charged particles, although a residual proton component remains within the selected window. Table 4 summarizes the fraction of events rejected by the upper threshold of 10 MeV for photons, neutrons, protons and the remaining secondary fragments. Photon rejection remains low (generally below about 5%), indicating that most photon induced events are preserved. In contrast, the threshold removes approximately 76-93% of secondary protons, depending on beam energy and detector position. A smaller fraction of secondary neutrons is also rejected, increasing from about 9-16% at detector position A to approximately 49-53% at the position D. Overall, these results indicate that the adopted energy selection effectively reduces charged particle contamination in the PGT analysis while retaining the majority of photon induced signals. The remaining proton contribution could be further suppressed in practical detector systems by implementing a charged particle veto by placing a thin plastic scintillator in front of the secondary radiation detector. Such a configuration would enable efficient discrimination of charged particles while preserving the prompt gamma signal.

Table 4 - Fraction of secondary particle rejected by energy deposition threshold above 10 MeV for different detector positions and beam energies. Percentages are calculated relative to the total number of events of each particle type reaching the detectors in the MC simulations.

| Beam Energy [MeV/u] | Detector Position | Gammas (%) | Neutrons (%) | Protons (%) | Other (%) |
|---|---|---|---|---|---|
| 166.41 | A | 1.07 | 9.14 | 76.49 | 1.86 |
|  | B | 1.38 | 21.45 | 81.72 | 5.28 |
|  | C | 2.27 | 35.07 | 90.28 | 75.04 |
|  | D | 2.73 | 49.26 | 92.02 | 92.14 |
| 268.86 | A | 0.98 | 12.76 | 82.03 | 5.80 |
|  | B | 1.33 | 26.81 | 88.96 | 7.02 |
|  | C | 2.51 | 39.97 | 92.14 | 33.80 |
|  | D | 3.86 | 52.25 | 93.03 | 91.70 |
| 398.84 | A | 4.60 | 15.94 | 86.69 | 50.25 |
|  | B | 3.16 | 31.38 | 90.00 | 42.22 |
|  | C | 3.49 | 44.35 | 92.24 | 32.96 |
|  | D | 5.44 | 53.40 | 93.01 | 77.66 |

*3.5 PG spatial distribution and production channels*

Figure 7 presents the spatial distribution of photon production reaching the secondary detectors at positions A, B, C, and D for 166.41, 268.86, and 398.84 MeV/u carbon ion beams. The prompt gamma component includes photons produced along the full interaction cascade, i.e., from interactions of the primary carbon ions as well as from secondary and tertiary fragments generated. MC simulations accurately reproduce the range of the carbon ion beams and help explain the variations observed in both MC and experimental distributions. For the 166.41 MeV/u carbon ion energy (Fig. 6), the height of the prompt gamma peak decreases from detector position A to D, dropping from approximately 4000 counts at position A to about 1200 counts at position D. For the 268.86 MeV/u beam (Fig. 11, Supplementary Materials), the prompt gamma peak remains nearly constant at around 4000 counts across detector positions, while for the 398.84 MeV/u beam (Fig. 12, Supplementary Materials), the peak increases from approximately 4000 counts at position A to about 12500 counts at position D. These trends are primarily determined by the relative position of the prompt gamma emission region with respect to the detectors, which is governed by the carbon ion range in the target. Because the detector positions are oriented toward the center of the PMMA target, variations in beam range modify both the effective detector solid angle subtended by the emission region and the photon path length through the target material (i.e., the absorption length), which affects photon attenuation.

For the 166.41 MeV/u, the range does not reach the middle of the target, so upstream detectors see more prompt gamma emission than downstream detectors, resulting in a decreasing peak. For the 268.86 MeV/u beam, the range reaches approximately the middle of the target, producing a roughly symmetric gamma emission across detector positions and nearly constant peak. For the 398.84 MeV beam, the range extends beyond the middle, so downstream detectors are closer to the high emission region, leading to an increasing peak from A to D. In all cases, stacking of secondary fragments, including neutrons and protons mainly, modulate the overall time distributions and contribute to the tail structure. This effect is significant compared with proton beams and reduces the suitability of PGT integration technique (Everaere et al. 2024). Overall, the production of prompt gamma rays is mostly localized along the beam axis, indicating the most probable interaction region between carbon ions and target nuclei, and explaining the correlation between the height of the PGT distributions and the carbon ion range. In addition, a smaller fraction of gamma rays appears to originate from the secondary detectors (Fig. 7, red outlines), possibly due to tertiary nuclear interactions and/or potential activation of the $LaBr_3(Ce)$ crystals (Cazzaniga et al. 2015, Ranga et al. 2024).

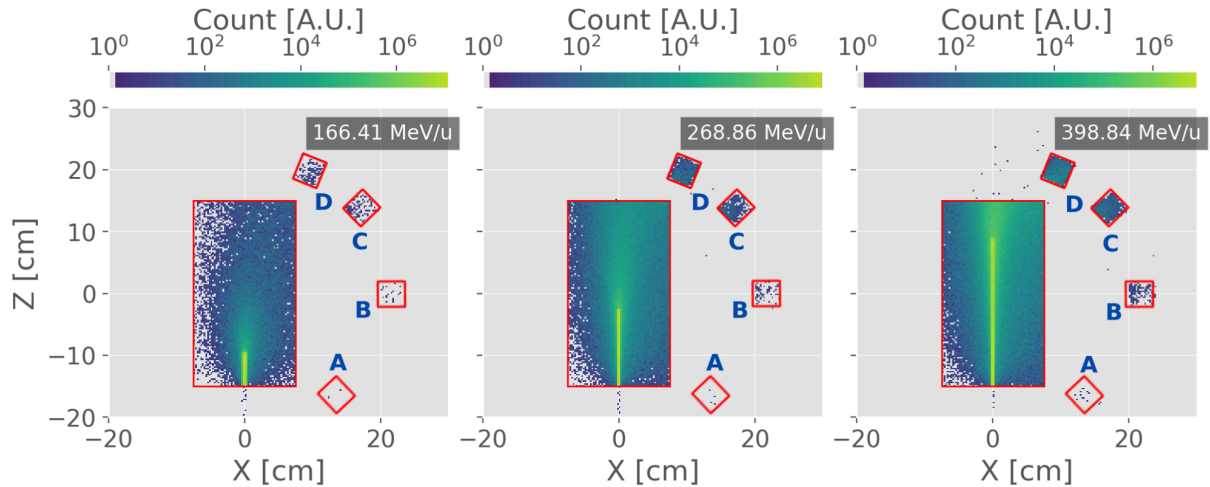

Figure 7 - MC spatial distribution of photon production for irradiations with 166.41 (left), 268.86 (middle), and 398.84 (right) MeV/u carbon ion beams. The maps include photons generated by interactions of primary carbon ions as well as those originating from secondary and tertiary interactions. Only photons depositing energy within within 1-10 MeV range in the secondary radiation detectors (outlined in red) are considered.

As a final analysis, the energy spectra of photons reaching the secondary detector positions are investigated. Figure 8A presents the distribution of energy deposited in the detectors as a function of the emitted photon energy for carbon ion simulations at 166.41, 268.86, and 398.84 MeV/u. Only events with deposited energy between 1 MeV and 10 MeV and within a time window up to 50 ns are considered, reproducing the selection criteria used in the PGT analysis. Photon scattering in the PMMA phantom and surrounding materials is fully taken into account in the MC simulations. Therefore, the deposited energy spectrum includes the effects of Compton scattering and other interaction processes that reduce the photon energy before reaching the detector. 511 keV annihilation photons are discarded by defaults.

As expected, multiple horizontal lines associated with prompt gamma production inside the target are visible. Lower intensity contributions arise from de-excitation of $^{11}$C* (2.00 MeV line), neutron capture on hydrogen (2.22 MeV line), $^{14}$N* (2.31 MeV line), $^{13}$C* (3.68 MeV and 3.85 MeV lines), $^{15}$O* (5.18 MeV line), $^{16}$O* (6.92 MeV and 7.12 MeV lines) (Verburg and Seco 2014, Wang et al. 2022, Freitas et al. 2025). The most prominent horizontal lines correspond to 4.44 MeV and 6.13 MeV gamma rays, originating from the de-excitation of $^{12}$C* and $^{16}$O*, respectively. These emissions result from interactions between carbon ions and the PMMA target, which are strongly correlated with the beam range (Golnik et al. 2014, Krimmer et al. 2018). Another relevant horizontal line corresponds to neutron capture on hydrogen, producing a tertiary prompt gamma at 2.22 MeV. Although this process is typically associated with delayed neutron capture, its contribution in the present spectra is reduced by the applied time selection. Table 5 reports the ratio of counts in the full-energy peaks to the corresponding Compton tails for the main prompt gamma lines detected at each carbon ion energy. Although the detector is primarily optimized for timing measurements rather than spectroscopy, these ratios highlight the contribution of Compton scattering in simulated spectra. The relatively low peak-to-Compton ratios reflect the limited detector size for detecting few-MeV photons and are consistent with the expected detector response.

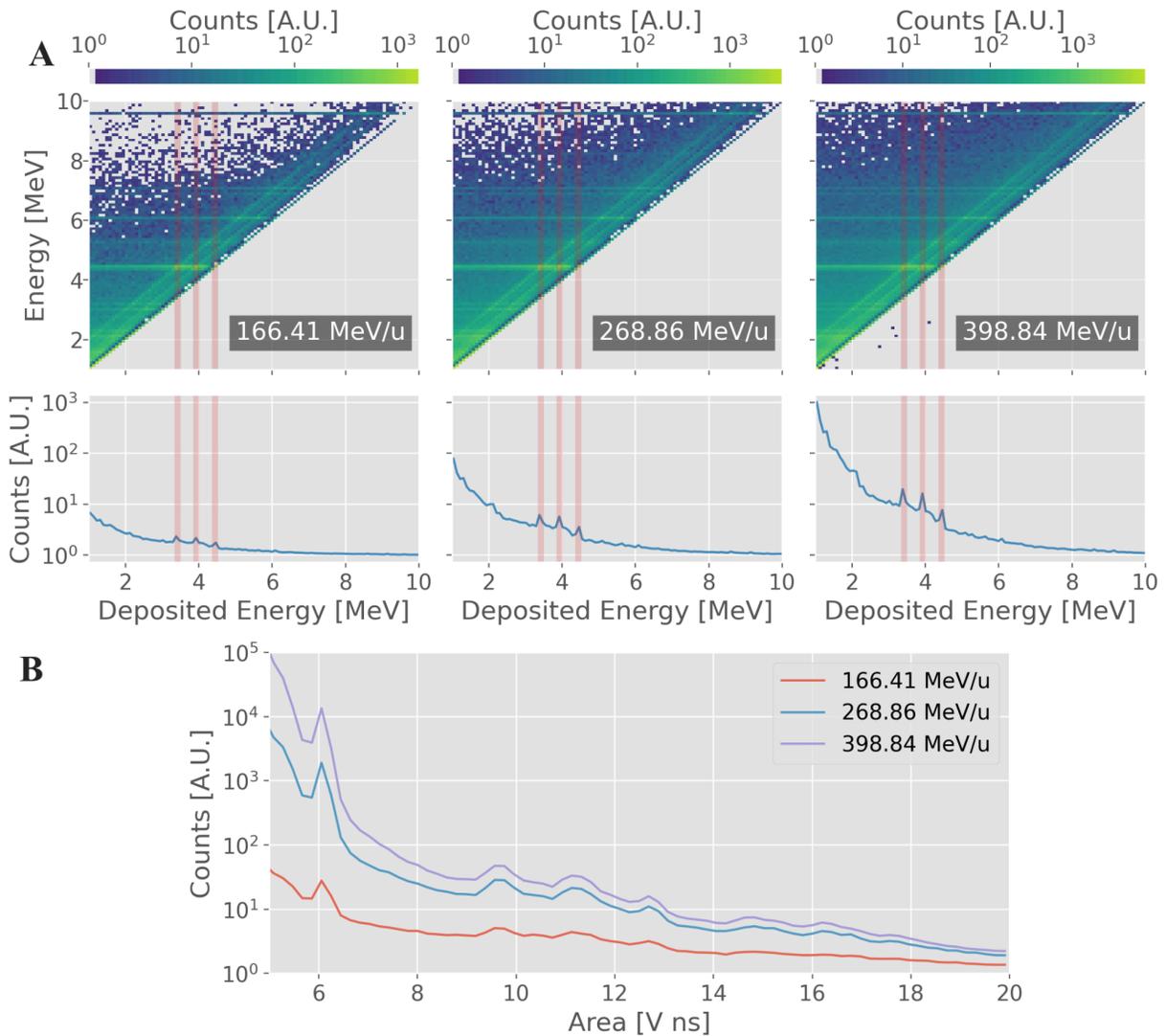

Figure 8 - (A) MC vs. deposited photon energy distribution during irradiation with 166.41 (left), 268.86 (middle), and 398.84 (right) MeV/u 398.84 MeV/u carbon ion beams. Cumulative energy distribution (bottoms). (B) Collected charge (signal area in V·ns) measured by the secondary radiation detector during the experimental data acquisitions for carbon ion irradiation at 166.41 MeV/u (purple), 268.86 MeV/u (blue), and 398.84 MeV/u (red). Distributions include all events recorded in the 1-10 MeV energy window across detector positions A-D.

Table 5 - Ratio of counts in the full energy peaks to the corresponding Compton tails for the main prompt gamma lines detected in MC simulations for each carbon ion energy.

| Beam Energy [MeV/u] | Photon Energy [MeV] | Peak/Compton ratio |
| --- | --- | --- |
| 166.41 | 2.22 | 0.2289 |
|  | 4.44 | 0.0805 |
|  | 6.13 | 0.0155 |
| 268.86 | 2.22 | 0.2301 |
|  | 4.44 | 0.0821 |
|  | 6.13 | 0.0163 |
| 398.84 | 2.22 | 0.2298 |
|  | 4.44 | 0.0800 |
|  | 6.13 | 0.0160 |

In addition, diagonal structures separated by 511 keV are also visible in Fig. 8A, indicating that pair production contributes significantly to the interaction mechanism in the $LaBr_3(Ce)$ crystals. The lowest diagonal corresponds to events in which the deposited energy equals the emitted photon energy, i.e., full energy deposition in the detector. Such events can arise from different interaction mechanisms, including photoelectric absorption and pair production followed by full absorption of the annihilation photons. The two upper diagonals are attributed to single-escape and double-escape events, respectively, where one or both 511 keV annihilation photons escape the crystal volume after a pair production interaction. Figure 8A also shows the energy deposition integrated over the photon energy axis for the three carbon ion energy cases. The three peaks highlighted by the red vertical lines indicate the system's capability to identify pair production interactions induced by 4.44 MeV photons within the crystals. This interpretation is further supported by Fig. 8B, which presents the distribution of collected charge measured by the secondary radiation detector during the experimental data acquisitions. The histograms are obtained by summing all events recorded in the 1-10 MeV energy window across the detector positions A, B, C, and D for each carbon ion beam energy, without applying combinatorial background rejection or time cuts to the PGT distributions. The total number of events contributing to each distribution is 1,240,587 for 166.41 MeV/u, 2,740,286 for 268.86 MeV/u, and 3,430,962 for 398.84 MeV/u. An additional early peak corresponding to a signal area around 6.0 V ns is also observed and may be associated with neutron capture interaction on hydrogen, followed by the emission of a 2.22 MeV prompt photon.

## 4 Conclusions

In this work, the first experimental results of online PGT monitoring of clinical-like treatment sessions with carbon ion beams of 166.41, 268.86, and 398.84 MeV/u kinetic energy irradiating a homogeneous 30.0 cm thick PMMA target are reported. Experimental data acquisitions were carried out using a dedicated detection system composed of a thin silicon strip sensor and a LaBr$_3$(Ce) scintillating crystal coupled to a SiPM matrix. The secondary radiation detector was placed in four different positions (labeled A, B, C, and D). A compatibility analysis was conducted between experimental and simulated PGT distributions, showing a good agreement between experimental and MC distributions. These results demonstrate the importance of MC simulations in interpreting PGT distributions for heavier ion irradiations and highlight how the detected distributions vary depending on detector setup. The study also shows that a bin-by-bin analysis is robust under statistics comparable to clinical scenarios. Future work will extend these analyses to inhomogeneous targets and explore the system sensitivity to smaller beam energy differences.

The particle contributions to the overall PGT signals were analysed for all detector positions and carbon ion energy combinations. Prompt gamma rays were identified as the dominant contribution at detector positions A and B, indicating that these positions provide the least signal contamination from neutrons and charged fragments. In contrast, detector positions C and D showed an increasing contribution from neutrons and charged fragments. Although these positions are more affected by such contamination, their contributions can be quantified using MC simulations. Furthermore, these findings indicate that the proposed setup may also be sensitive to neutron-induced signals. Although prompt gamma detection remains the primary objective for beam range verification, the capability to observe neutron-related events could provide complementary information on nuclear fragmentation processes and on the characterization of secondary radiation fields, including neutrons that contribute to the out-of-field dose in particle therapy.

The spatial distribution of gamma rays reaching the detector positions was studied for all carbon ion beams. MC simulations reproduce the expected spatial correlation between photon production and the beam trajectory, consistent with the physical basis of the PGT technique. The detected signal is predominantly produced by prompt photons generated in nuclear interactions between the carbon ions and the target nuclei, while the observed spectral features arise from the standard interaction processes of these photons within the scintillator (e.g., Compton scattering and pair production).

The energy deposition analysis revealed a prominent contribution of prompt gammas arising from the de-excitation of carbon (4.44 MeV) and oxygen (6.13 MeV) after direct interactions with carbon ions. MC simulations indicate that 4.44 MeV photons interact in the scintillator through both Compton scattering and pair production, with the latter producing characteristic spectral features such as single- and double-escape events. These features are consistent with the structures observed in the collected charge distribution from the experimental data acquisitions. Additional prompt gamma rays not directly correlated with the primary particle trajectory were also observed, although they contributed less to the overall signal. Furthermore, prompt gamma production following neutron capture within the target was identified, suggesting a potential indirect method for assessing out-of-field dose related to neutron production in beam-target fragmentation processes.

## Acknowledgement

The authors thank the INFN electronic and mechanical service staff for their assistance in the preparation of the experimental setup, and the CNAO personnel for their technical support and provision of beamtime.

## Funding

This study was funded by the SIG (Superconducting Ion Gantry) project (CUP:I55F21003740001), funded by the National Institute of Nuclear Physics (INFN, Call CSN5 2021) under the Italian National Research Program (PNR), funded by the Italian Ministry of University and Research (MUR). This study was also funded by the HONEY (Hybrid ONline tEchnology for particle therapy) project (CUP: I53D23001280006), funded within the PRIN 2022 program by the Italian Ministry of University and Research (MUR).

## Author Contributions

I.B.R.O. and S.R. contributed equally to this work. I.B.R.O. led the Monte Carlo simulation framework, data analysis and manuscript writing. S.R. and E.F. led the experimental measurements and data analysis. S.R., P.C., E.M.D., M.F.M., M.D.F.M., V.F., S.G., F.M.M., D.M.M.O., F.P. and E.F. provided experimental and instrumental support, developed and characterised the detectors and performed the experimental measurements. M.P. and S.S. provided technical expertise, CNAO facility access and beam time. P.C., R.S., R.C., A.V., and E.F. provided senior supervision and project coordination. All authors contributed to manuscript preparation and approved the final version.

## Data Availability

The data supporting the findings of this study are available from the corresponding author upon reasonable request.

## Supplementary Materials

Supplementary material containing results for carbon ion beams at 268.86 MeV/u and 398.84 MeV/u kinetic energies. Figure 9 and 10 present the experimental and MC simulated time distributions for all detector positions at these higher energies, with bin-by-bin compatibility tests showing good agreement between measurements and simulations. Figure 11 and 12 display the particle contributions to the PGT distributions, illustrating the evolution of secondary particle (neutrons, protons, electrons, positrons) contributions as a function of beam energy and detector position. These results corroborate the findings from the main text and extend the validation of the PGT technique and Geant4 simulation framework to higher energies relevant for clinical carbon ion therapy.

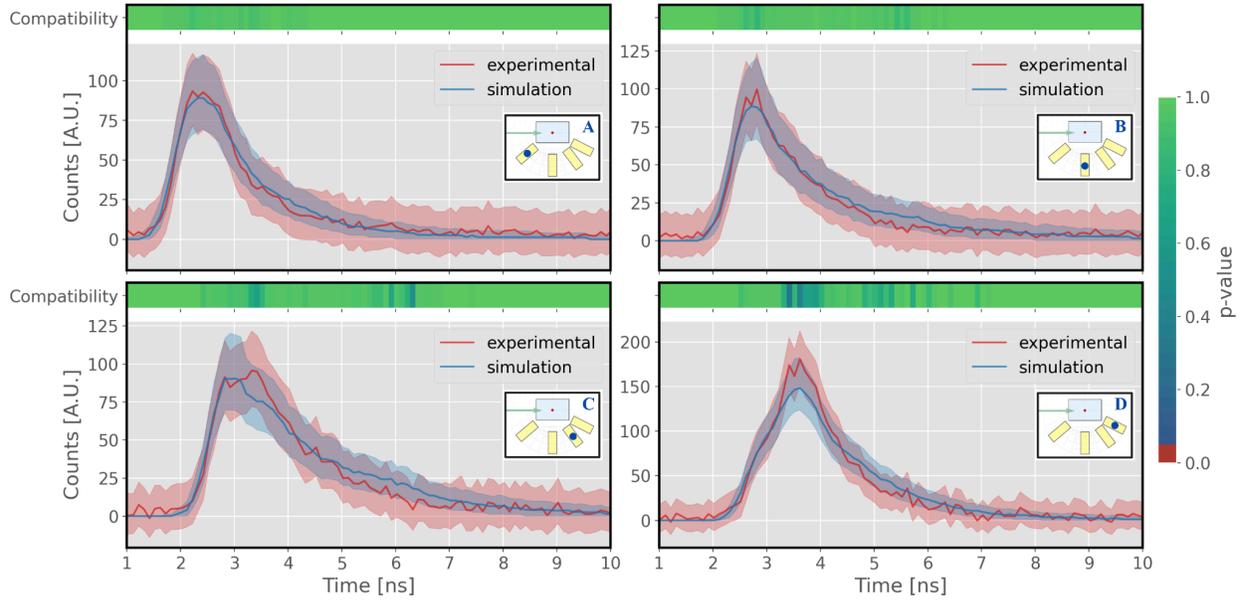

Figure 9 - Experimental and MC time distributions for a 268.86 MeV/u carbon ion beam at each detector position (shown in the box in each subplot). For each position of the secondary radiation detector, compatibility test (top) and PGT distribution histograms with 95% CIs (bottom).

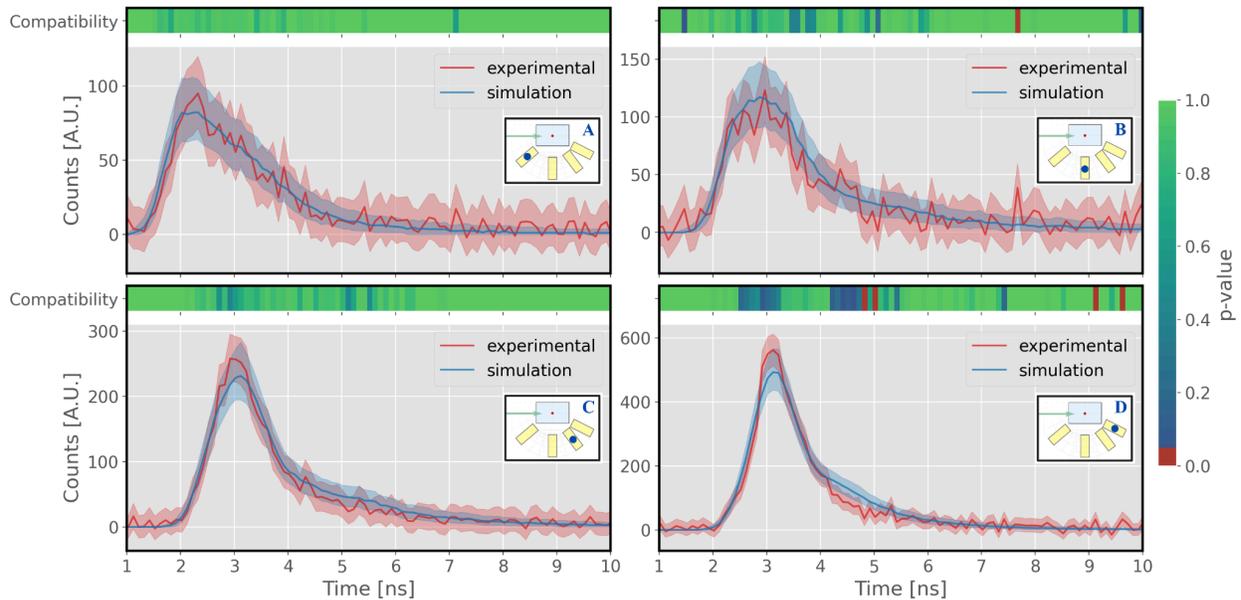

Figure 10 - Experimental and MC time distributions for a 398.84 MeV/u carbon ion beam at each detector position (shown in the box in each subplot). For each position of the secondary radiation detector, compatibility test (top) and PGT distribution histograms with 95% CIs (bottom).

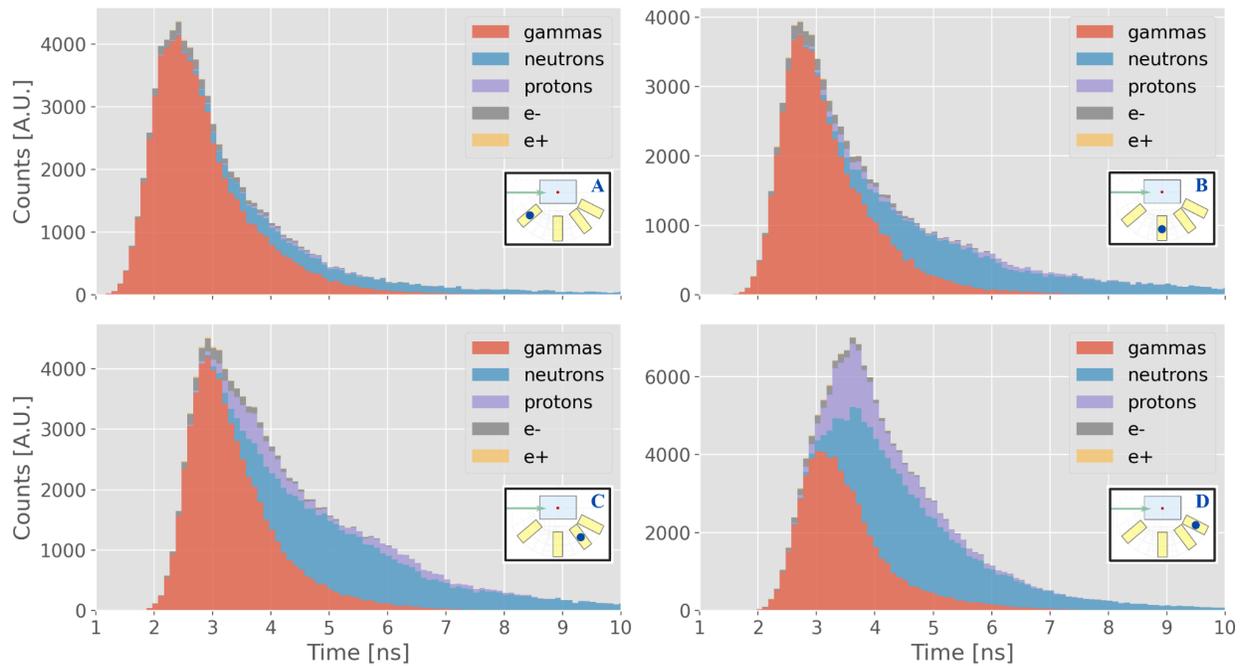

Figure 11 - Particle contribution to the PGT distributions for a 268.86 MeV/u carbon ion beam at detector position A (top left), B (top right), C (bottom left), and D (bottom right).

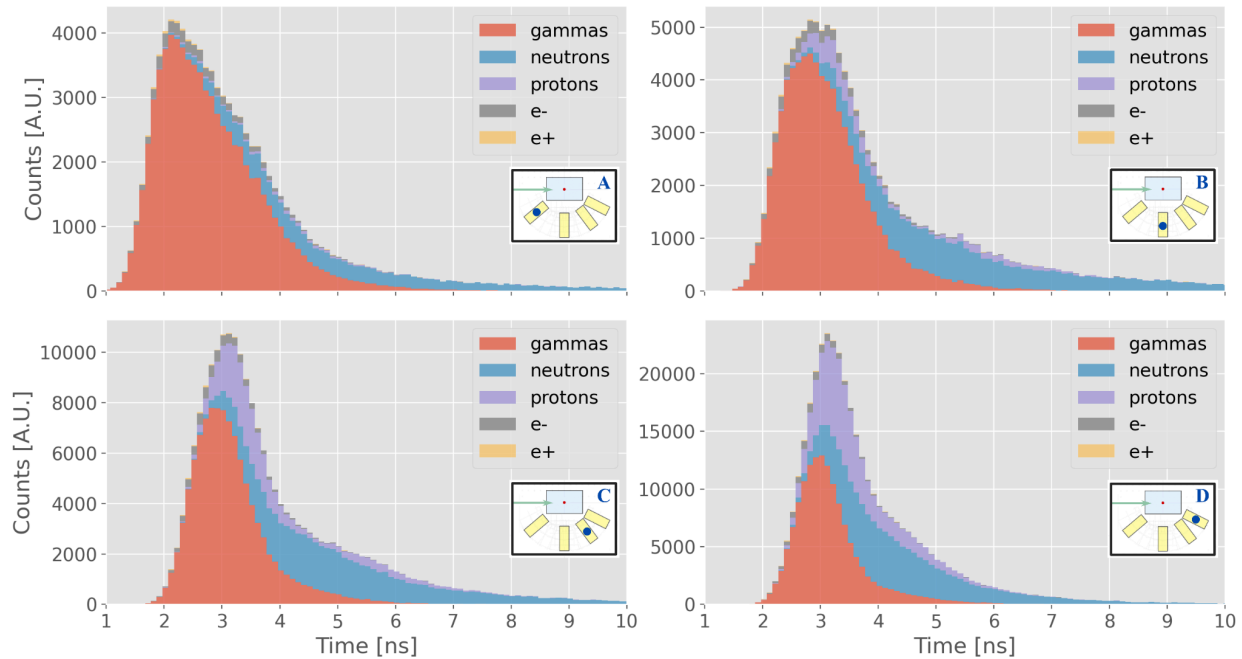

Figure 12 - Particle contribution to the PGT distributions for a 398.84 MeV/u carbon ion beam at detector position A (top left), B (top right), C (bottom left), and D (bottom right).